\documentclass[twocolumn,showpacs,preprintnumbers,amsmath,amssymb]{revtex4}

\usepackage{graphicx}
\usepackage{dcolumn}
\usepackage{bm}
\usepackage{color}

\begin{document}

\title{Deterministic Vector Freak Waves}

%
%
%
%
%

\author{Fabio Baronio$^1$, Antonio Degasperis$^2$, Matteo Conforti$^1$, and Stefan Wabnitz$^1$}

\affiliation{$^1$CNISM, Dipartimento di Ingegneria dell'Informazione, Universit\`a di Brescia, Via Branze 38, Brescia 25123, Italy.\\
$^2$ INFN, ``Sapienza'' Universit\`a di Roma, P.le A. Moro 2, 00185 Roma, Italy}
\date{\today}

\begin{abstract}
We construct and discuss a semi-rational, multi-parametric vector solution of coupled nonlinear Schr\"odinger equations (Manakov system). This family of solutions includes known vector Peregrine solutions, bright-dark-rogue 
solutions, and novel vector unusual freak waves. The vector freak (or rogue) waves could be of great interest
in a variety of complex systems, from optics to Bose-Einstein condensates and finance.  
\end{abstract}

\maketitle

\textit{Introduction.}
Extreme wave events, also referred to as freak or rogue
waves, are mostly known as an oceanic phenomenon responsible
for a large number of maritime disasters. These
waves, which have height and steepness much greater
than expected from the sea average state \cite{dhyste2008}, have recently become a topic
of intense research. 
Freak waves appear both in deep ocean and in
shallow waters \cite{akhmediev2010sp}. In contrast to tsunamis and 
storms associated with typhoons that can be predicted hours 
(sometimes days) in advance, the particular danger of oceanic rogue 
waves is that they suddenly appear from nowhere only 
seconds before they hit a ship. The grim reality, however, is that
although the existence of freak waves has now been confirmed by multiple observations, 
uncertainty remains on their fundamental
origins. This hinders systematic approaches to study their characteristics,
including the predictability of their appearance \cite{pelinosky2008}.

In fact, research on rogue waves is in an emerging state
\cite{dhyste2008,pelinosky2008,osborne2010}. These waves
not only appear in oceans \cite{garret2009} but 
also in the atmosphere \cite{stenflo2009}, in optics \cite{solli2007,erkin2009},
in plasmas \cite{moslem2011}, in superfluids, in Bose–-Einstein condensates \cite{bludov2009} and also
as capillary waves \cite{xia2010}. The common features and differences among freak wave manifestations in their different contexts is a subject of intense discussion \cite{akhmediev2010sp}.
New studies of rogue waves in any of these disciplines enrich
their concept and lead to progress towards a comprehensive understanding
of a phenomenon which still remains largely unexplored.
A formal mathematical description of a rogue wave is provided by the 
so-called Peregrine soliton \cite{peregrine83}. This solitary wave is a solution of the
one-dimensional nonlinear Schr\"odinger equation (NLSE) with the property of being localized in both the transverse and 
the longitudinal coordinate: thus it describes a unique wave event. This solution is also
unique in a mathematical sense, as it is written in terms of rational
functions of coordinates, in contrast to most of other known solutions of the
NLSE. Recent experiments have provided a path to 
generating Peregrine solitons in optical fibers with standard telecommunication equipment \cite{kibler2010}. The further experimental observation of Peregrine solitons
in a water tank \cite{chab2011} indicates that they can also describe
rogue waves in oceans.
The Peregrine soliton is not the only fully localized waveform \cite{akhmediev2009pla}. 
In fact, there is an infinite hierarchy of rational solutions of the NLSE
which enjoy the same property \cite{akhmediev2009pre,matveev2010,gaillard2011,guo12}.

In a variety of complex systems such as Bose--Einstein condensates \cite{akhmediev10epj}, 
optical fibers \cite{kaup93}, and financial 
systems \cite{ivancevic2009,yan2011}, several variables rather than a single wave 
amplitude need to be considered. For instance, in the financial world it is necessary to couple cash to the value of other assets 
such as shares, bonds, options, etc., as well as to consider all correlations 
between these variables. The resulting systems of equations may thus describe extreme waves with 
higher accuracy than the single NLSE model. Approaches to rogue wave phenomena 
involving multiple coupled waves are the coupled Gross-Pitaevskii (GP) equations \cite{pita2003} and the Manakov system (or vector NLSE) \cite{manakov74}.
Indeed, vector rogue waves of GP equations and the Manakov system have been recently presented 
\cite{akhmediev10epj,yan2011,guo2011}.

In this Letter, we construct and discuss a novel \emph{semi-rational}, multi-parametric vector solution of the 
Manakov system. For special parameter values this solution reproduces known vector rogue waves 
(such as the vector Peregrine soliton \cite{akhmediev10epj} and bright-dark-rogue waves \cite{guo2011}). 
Our treatment below goes as follows. We give the essential Darboux dressing 
transformation to construct freak solutions of the Manakov system. 
We present the multi-parametric, semi-rational deterministic freak waves. 
Moreover, we discuss their experimental feasibility in nonlinear optics.

\textit{Darboux dressing technique.}\label{sec2}
 Waves are assumed to be modeled by the dimensionless vector coupled nonlinear Schr\"odinger equations (VNLSE) or Manakov system:

\begin{equation}\label{VNLS}
\left \{ \begin{array} {lll}
iu^{(1)}_t+ u^{(1)}_{xx} +2 ( |u^{(1)}|^2 +  |u^{(2)}|^2 ) u^{(1)} & = & 0 \\
iu^{(2)}_t+ u^{(2)}_{xx} +2 ( |u^{(1)}|^2  + |u^{(2)}|^2 ) u^{(2)} & = & 0, 
\end{array} \right .
\end{equation}
where $u^{(1)}(x,t),\,u^{(2)}(x,t)$ represent the wave envelopes and $t,x$ are the longitudinal and 
transverse coordinates, respectively. Each subscripted variable in Eqs. (\ref{VNLS}) stands for partial differentiation. 
It should be pointed out that the meaning of the dependent variables $u^{(1)}(x,t),u^{(2)}(x,t)$, and of the coordinates $t,x$ depends on the particular applicative context (f.i. plasma physics, nonlinear optics, finance). Note also that Eqs. (\ref{VNLS}) refer to the self-focusing (or anomalous dispersion) regime. 
Eqs. (\ref{VNLS}) are integrable: the associated Lax pair is
\begin{equation}\label{lax}
\Psi_x=(ik\sigma+Q)\Psi\;,\;\Psi_t=[2ik^2\sigma+2kQ +i\sigma(Q^2-Q_x)]\Psi,
\end{equation}
where $\Psi=\Psi(x,t,k)$ is a $ 3\times 3$ matrix solution, $k$ is the complex spectral variable, the matrix  $\sigma=\text{diag}\{1\,,\,-1\,,\,-1\}$ is constant and diagonal,
 and $Q=Q(x,t)$ is the $ 3\times 3$ matrix
 \begin{equation}\label{Q}
Q=\left ( \begin{array} {ccc}
0 & -u^{(1)*} & -u^{(2)*} \\ u^{(1)} & 0 & 0 \\ u^{(2)} & 0 & 0 \end{array} \right ).
\end{equation}
The starting point here is the construction of the solution representing one soliton wave nonlinearly superimposed to the following plane wave background solution of Eqs. (\ref{VNLS})
\begin{equation}\label{background}
\left ( \begin{array} {l}
u^{(1)}_0(x,t) \\ u^{(2)}_0(x,t) \end{array} \right ) = e^{2i\omega t}  \left ( \begin{array} {l}
a_{1} \\ a_{2} \end{array} \right ),
\end{equation} 
where $a_1$ and $a_2$ are arbitrary parameters which, with no loss of generality, are taken real. Moreover the frequency $\omega$ reads as $\omega=a^2$ where, from now on, we set $a=\sqrt{a_1^2+a_2^2}$.  
The Darboux method to obtain such one--soliton solution $u^{(1)}(x,t),\,u^{(2)}(x,t)$ is well known, therefore we omit detailed computations, limiting ourselves to list the essential few steps.  Our results have been obtained by following the general formulation and construction as presented in \cite{DL07} (the interested reader may find additional references quoted there).  

The chain of calculations ends up with the following general formula \cite{DL07}
\begin{equation}\label{1soliton}
\left ( \begin{array} {l}
u^{(1)} \\ u^{(2)} \end{array} \right ) = e^{2i\omega t} \left ( \begin{array} {l}
a_{1} \\ a_{2} \end{array} \right ) + \frac{2i(\beta^*-\beta) \zeta^*}{|\zeta|^2+z^{\dagger}z} \left ( \begin{array} {r}
z^{(1)} \\ z^{(2)} \end{array} \right ).
\end{equation}
The constant parameter $\beta$ is complex (with non vanishing imaginary part), while the functions $\zeta(x,t)\,,\,z^{(1)}(x,t)\,,\,z^{(2)}(x,t)$ are the components of a generic 3--dimensional vector solution $Z(x,t)$ of the Lax pair of equations (\ref{lax}), which corresponds to the spectral parameter $k=\beta$ and to the background solution (\ref{background}). Thus, if $Z_0$ is an arbitrary complex 3--dimensional vector, $Z(x,t)$ reads as
\begin{equation}\label{zvector}
 Z= \left ( \begin{array} {c}
\zeta  \\ z^{(1)} \\ z^{(2)} \end{array} \right )=\left ( \begin{array} {ccc}
1 & 0 & 0 \\ 0 & e^{2i \omega t} & 0 \\ 0 & 0 & e^{2i\omega t} \end{array} \right )\exp(i\Lambda x - i \Omega t) Z_0,
\end{equation}
where $\Lambda$ and $\Omega$ are the constant matrices
 \begin{equation}\label{lambdaomega}
\Lambda=\left ( \begin{array} {ccc}
\beta & ia_{1} & ia_{2} \\ -ia_{1} &  -\beta & 0 \\ -ia_{2} & 0 & -\beta \end{array} \right ), \; \;
\Omega= -\Lambda ^2 -2\beta \Lambda +\beta^2 + 2a^2.
\end{equation}
Formula (\ref{zvector}) shows that, if the matrix $\Lambda$ (and therefore $\Omega$) possesses three linearly independent eigenvectors, then the vector $Z(x,t)$ is a linear combination of exponential functions of $x$ and $t$. Therefore the solution (\ref{1soliton}) cannot be rational or semi--rational. Since only the exponential function of a nilpotent matrix is  polynomial, one has to find those particular values of $\beta$ (see (\ref{lambdaomega})) such that the matrix $\Lambda$ (and therefore $\Omega$) is similar to a Jordan matrix. Indeed this happens if and only if  $\beta=\pm ia$. By taking f.i. $\beta=ia$, in this way we arrive to the following semi--rational solution of the VNLSE Eqs. (\ref{VNLS})
\begin{equation}\label{soliton}
 \left ( \begin{array} {l}
u^{(1)}(x,t) \\ u^{(2)}(x,t) \end{array} \right ) = e^{2i\omega t} \left[ \frac{L}{B} \left ( \begin{array} {l}
a_{1} \\ a_{2} \end{array} \right ) + \frac{M}{B} \left ( \begin{array} {r}
a_{2} \\ - a_{1} \end{array} \right ) \right ],
\end{equation}
with the following notation:
$ L = \frac32 -8\omega^2 t^2 -2 a^2 x^2 +8i\omega t + |f|^2 e^{2a x}\;,  $
$ M= 4f (a x -2i\omega t -\frac 12 ) e^{(a x + i \omega t)}\;, $ and
$ B= \frac12 +8\omega^2 t^2 + 2 a^2 x^2 + |f|^2 e^{2a x} $, where
$ f $ is a complex arbitrary constant. 
It should be remarked that the dressing construction of the vector rogue wave (\ref{soliton}) 
has introduced as arbitrary parameters the three complex components of the vector $Z_0$, see  
(\ref{1soliton}) and (\ref{zvector}). However only the complex parameter $f$ is left essential 
out of these components, since the other parameters enter as trivial translations of the 
coordinates $x\,,\,t$, which have been set to zero for simplicity. The two other
 real parameters $a_1, a_2$ originate instead from the naked solution, namely
 from the background plane wave (\ref{background}). We note also that the dependence 
 of $L, M$ and $B$ (see (\ref {soliton})) on $x,t$ is both polynomial and exponential 
 only through the dimensionless variables $ax$ and $\omega t=a^2 t$. Moreover the vector 
 solution (\ref {soliton}) turns out to be a combination of the two constant orthogonal vectors $(a_1\,,\,a_2)^T$ and $(a_2\,,\,-a_1)^T$. 
 
\textit{Vector semi--rational rogue waves.}\label{sec3T}
Setting $f=0$ implies $M=0$: in this particular case the expression (\ref{soliton}) 
yields the trivial vector generalization of the 
rational Peregrine solution \cite{peregrine83, akhmediev10epj}. 
In this case $u^{(1)}(x,t)$ is merely proportional to $u^{(2)}(x,t)$. 
For future reference we note that the amplitude $|u^{(j)}(x,t)|$ is picked at 
$x=0$ with the maximum value $3 |a_j|$ at $t=0$. 

\begin{figure}[h]
\begin{center}
\includegraphics[width=4cm]{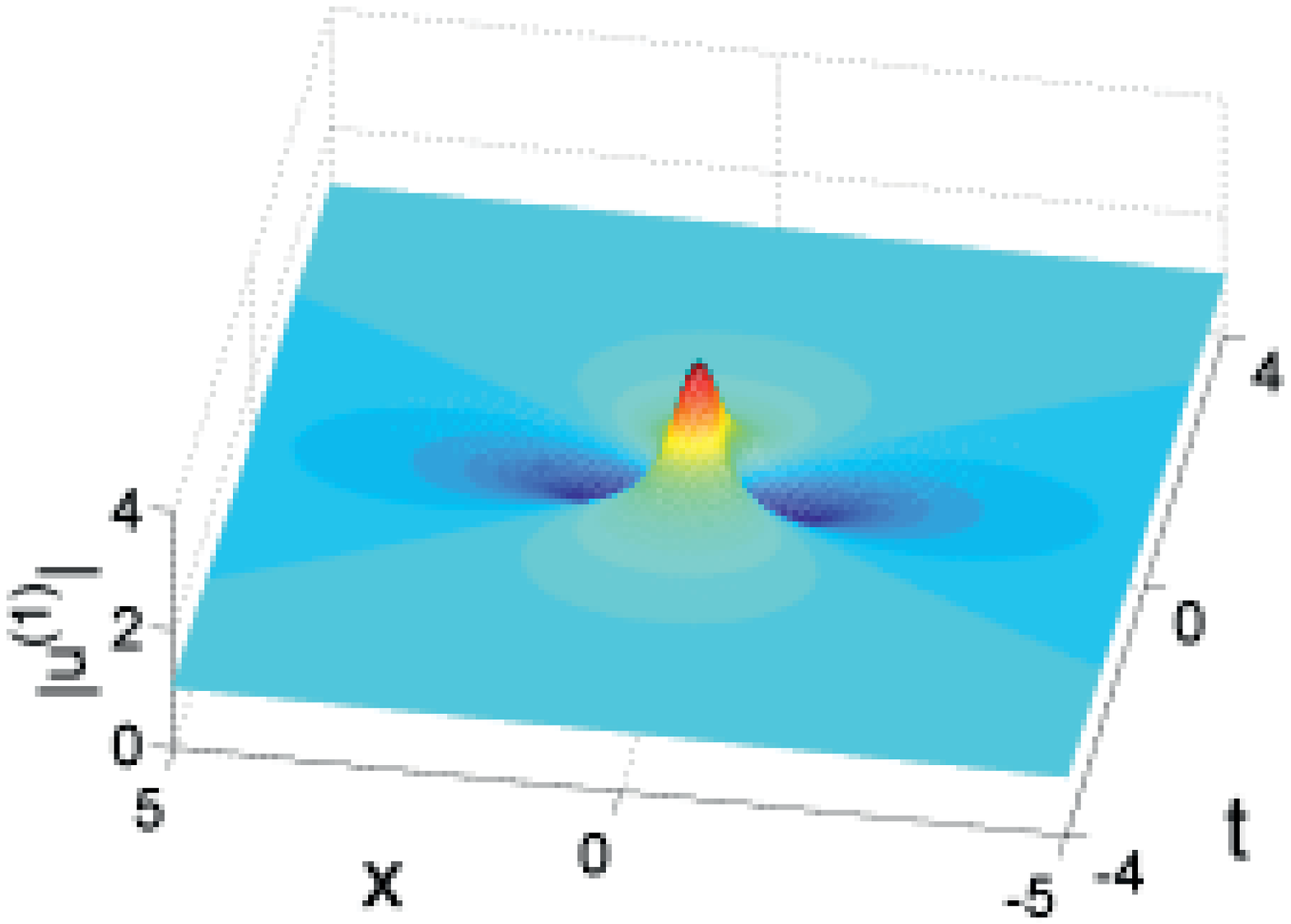}
\includegraphics[width=4cm]{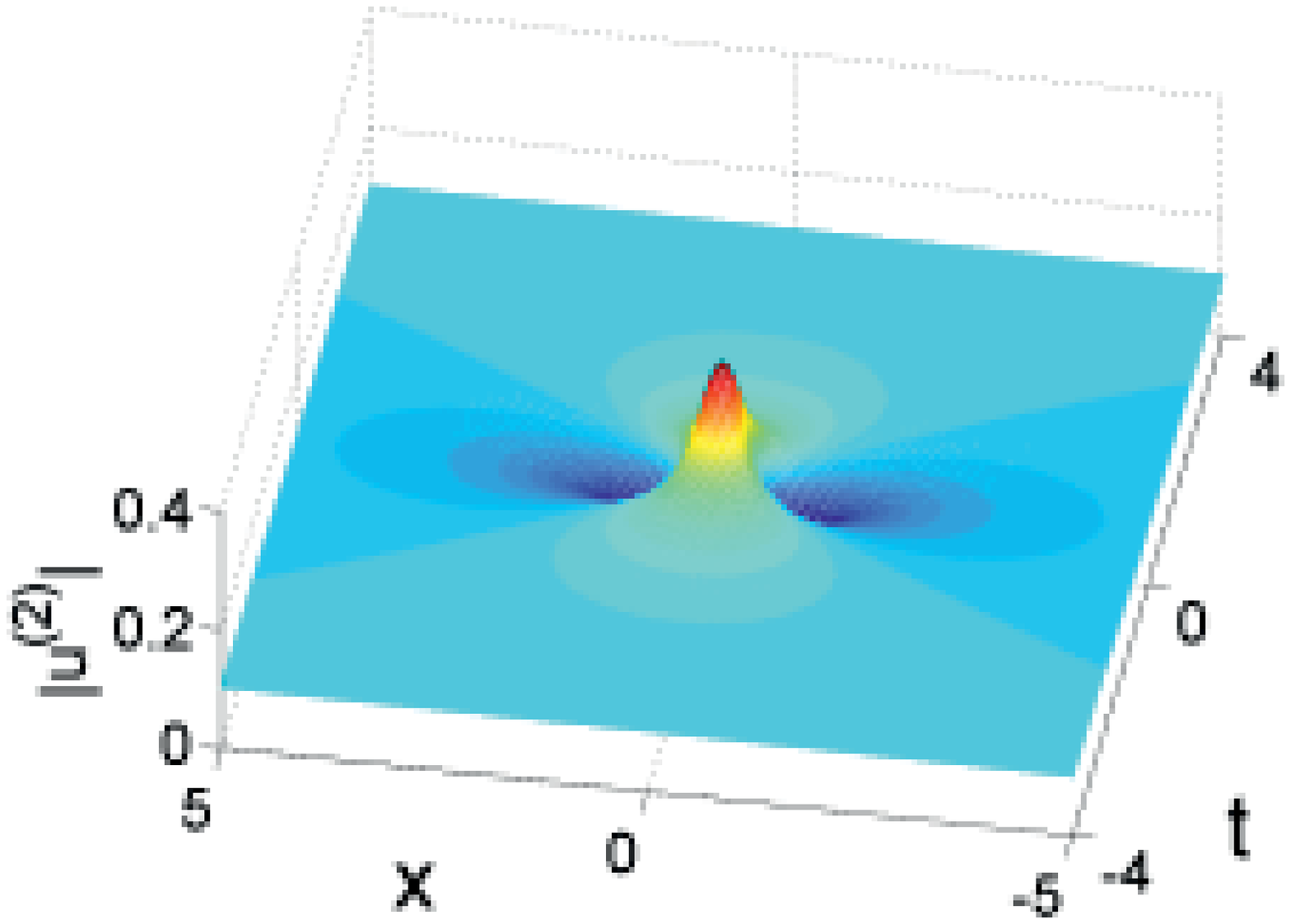}
    \end{center}
     \caption{Deterministic vector freak waves envelope distributions $|u^{(1)}(x,t)|$
     and $|u^{(2)}(x,t)|$ of (\ref{soliton}). Here, $a_1=1, a_2=0.1, f=0$.
    } \label{fig_peregrine}
\end{figure}

If instead $f\neq 0$, the Peregrine bump coexists and interacts with a pulse which 
propagates with non constant speed, and, depending on the value of $|f|$, 
may have different looks. In order to better describe this behavior we note that the ratios 
$L(x,t)/B(x,t)$ and $M(x,t)/B(x,t)$ which appear in (\ref{soliton}) describe asymptotically 
 as $t\rightarrow \pm \infty$ a dark and a bright pulse, respectively. Thus each wave 
 component $u^{(j)}(x,t)$ is generically a mixture of a dark and a bright pulse. Leaving 
 aside the detailed analytic description of the solution (\ref{soliton}) at intermediate 
 times, we limit our present analysis to the large time behavior. The pulse motion, for each 
 individual dark and bright contribution, asymptotically reads as

\begin{equation}\label{motion}
x=\xi(t) \rightarrow x_0 +\frac1a \ln(\omega |t|) + O\left(\frac{\ln^2(|t|)}{t^2}\right),\,\,t\rightarrow \pm \infty,
\end{equation}

where $x_0=(1/a) \ln(2\sqrt{2}/|f|)$. This implies that this pulse goes to infinity where it 
``stops'' since its velocity slowly vanishes as $d\xi/dt \rightarrow 1/(at)$.
 The shape of the dark and bright contributions at large times as a function of the parameter 
 $\chi$ which measures the displacement from the peak position, takes the expected form as 
 $t\rightarrow \pm \infty$

\begin{equation}\label{pulse}
 \frac LB \rightarrow \tanh(\chi) \,,\;\frac MB \rightarrow -i\sqrt{2}\left(\frac{f t}{|f t|}\right) \frac{e^{i\omega t}}{\cosh(\chi)} \;.
\end{equation}

\begin{figure}[h]
\begin{center}
\includegraphics[width=4cm]{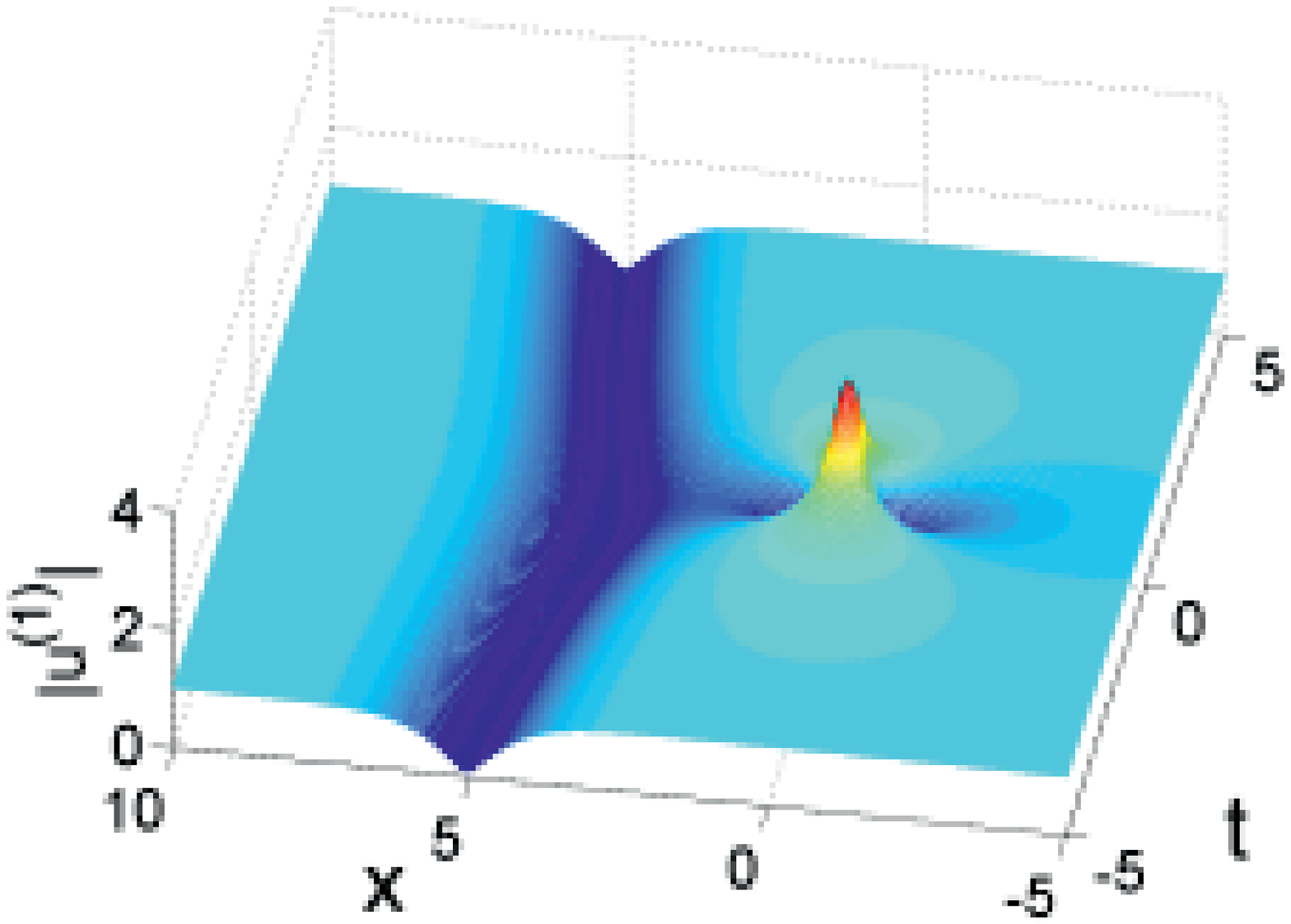}
\includegraphics[width=4cm]{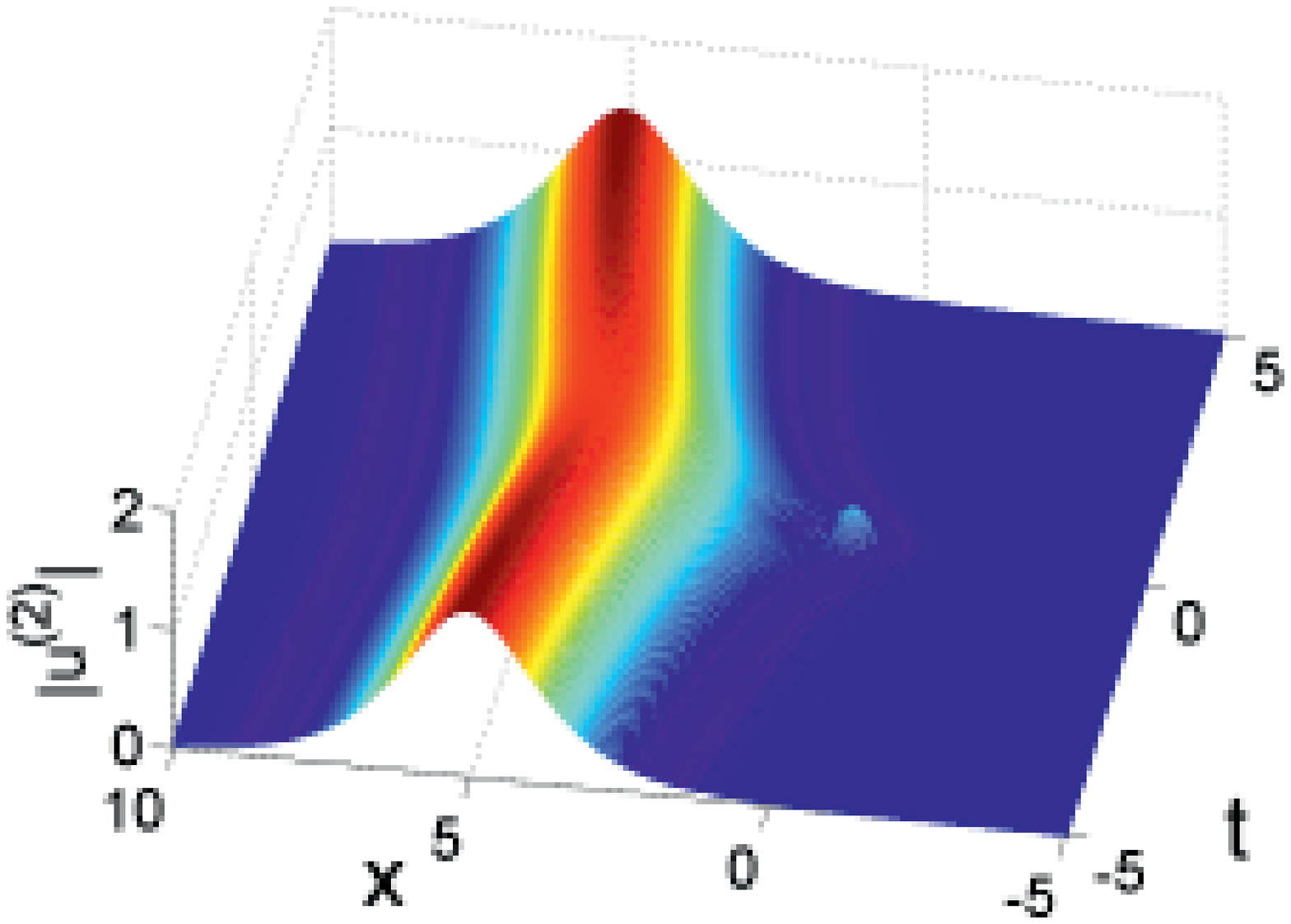}
    \end{center}
     \caption{As in Fig. \ref{fig_peregrine}, with $f=0.1, a_1=1, a_2=0$.
    } \label{fig_bdr1}
\end{figure}

\begin{figure}[h]
\begin{center}
\includegraphics[width=4cm]{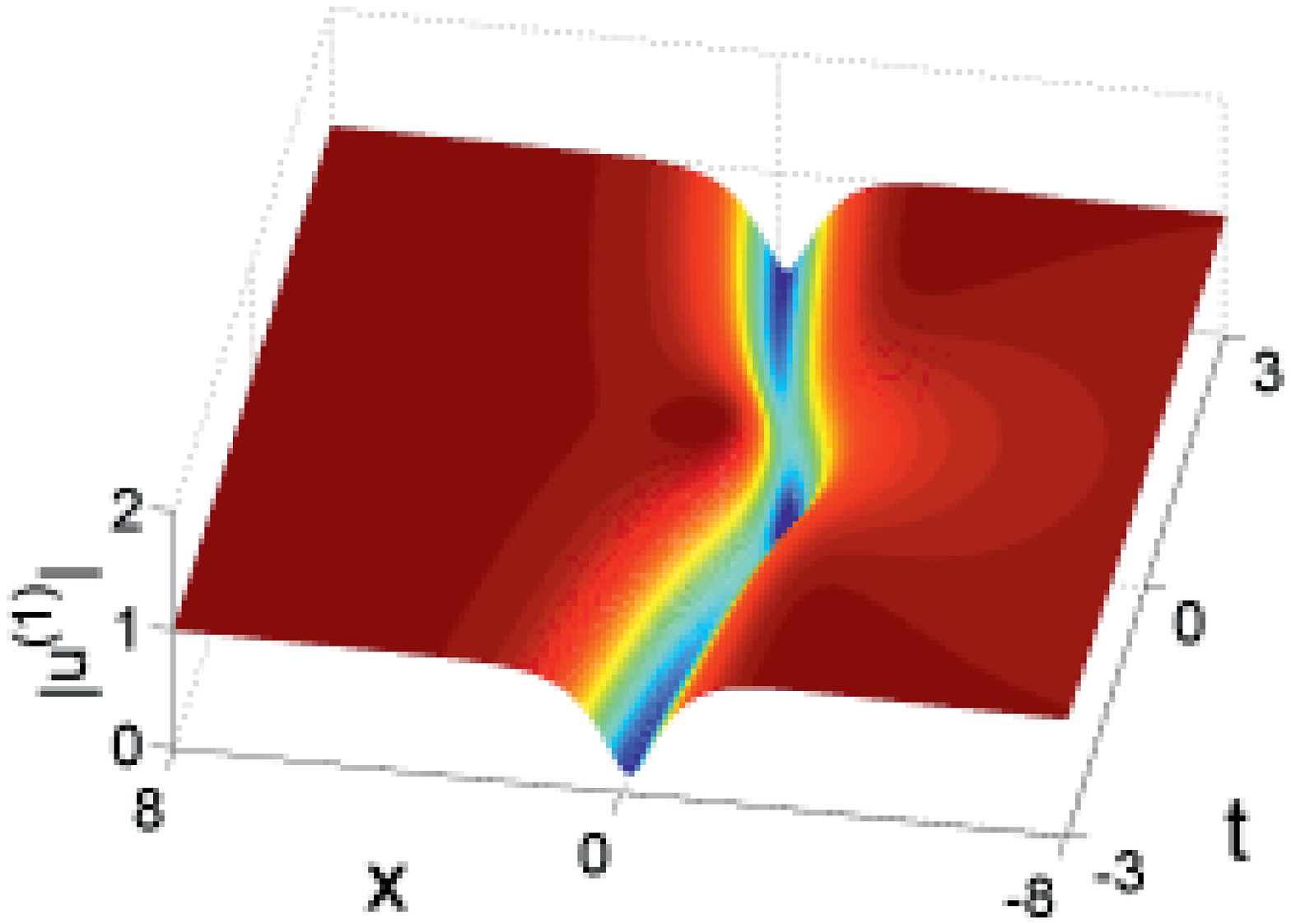}
\includegraphics[width=4cm]{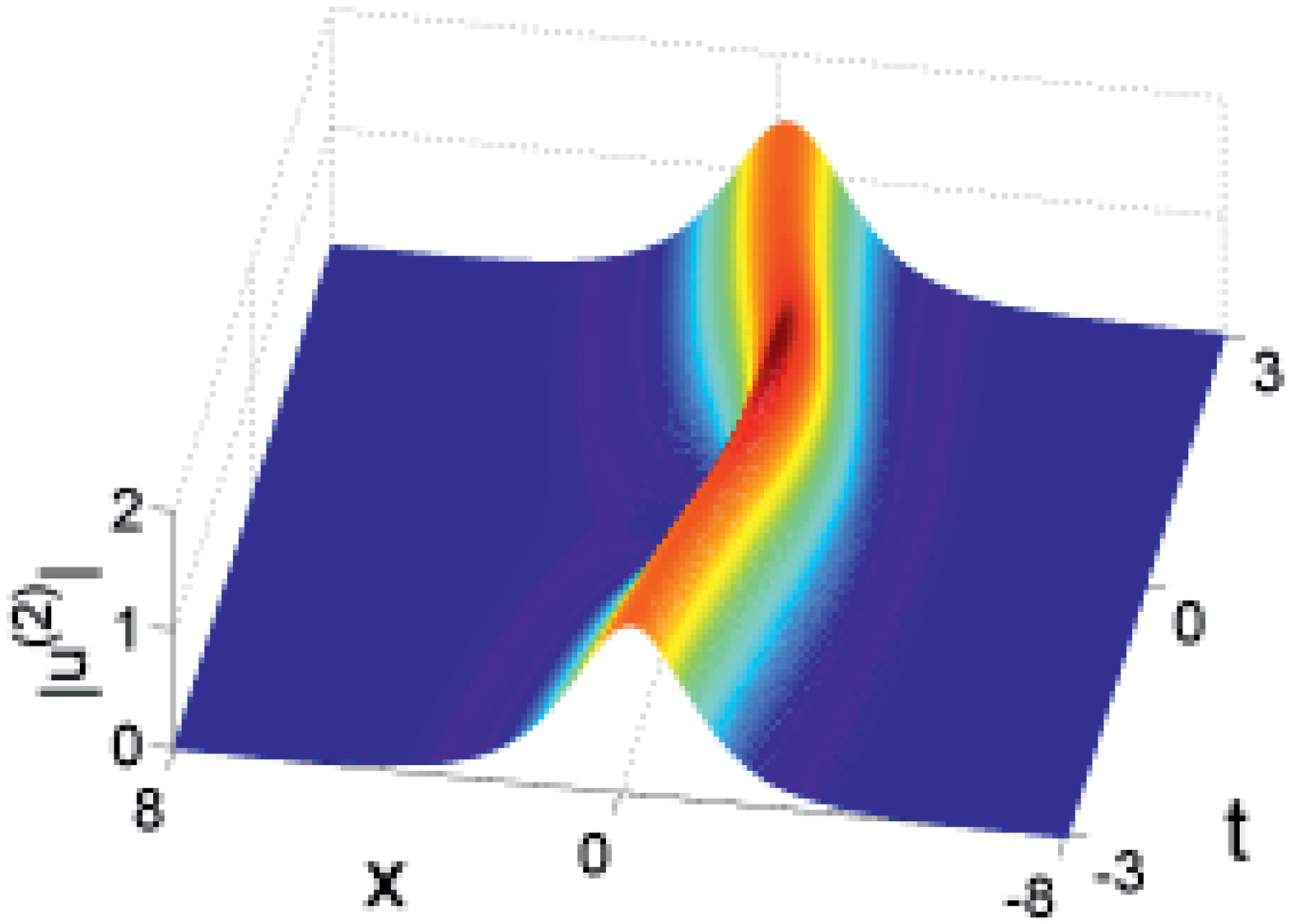}
    \end{center}
     \caption{As in Fig. \ref{fig_peregrine}, with $f=10, a_1=1, a_2=0$.
    } \label{fig_boom}
\end{figure}

\begin{figure}[h]
\begin{center}
\includegraphics[width=4cm]{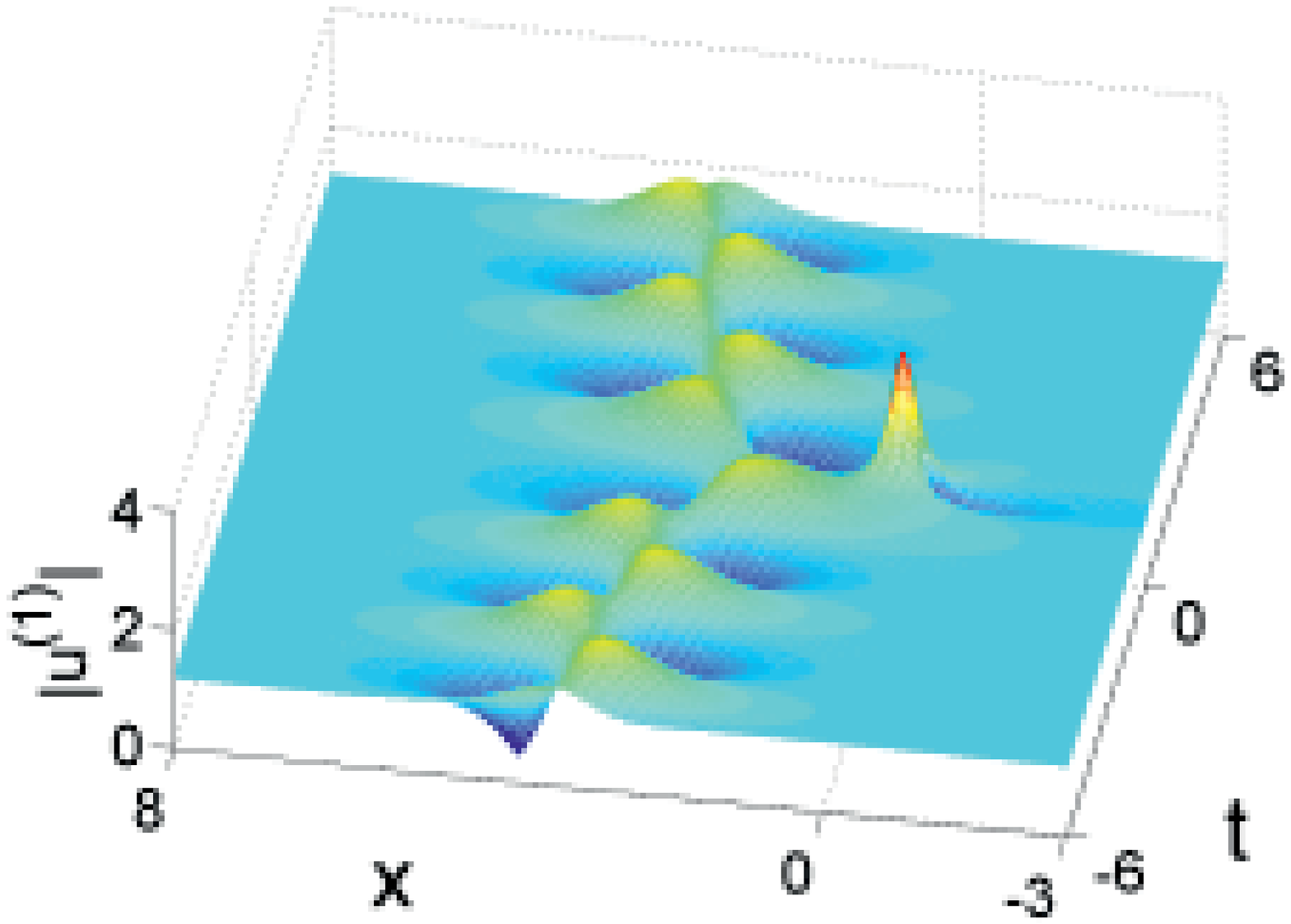}
\includegraphics[width=4cm]{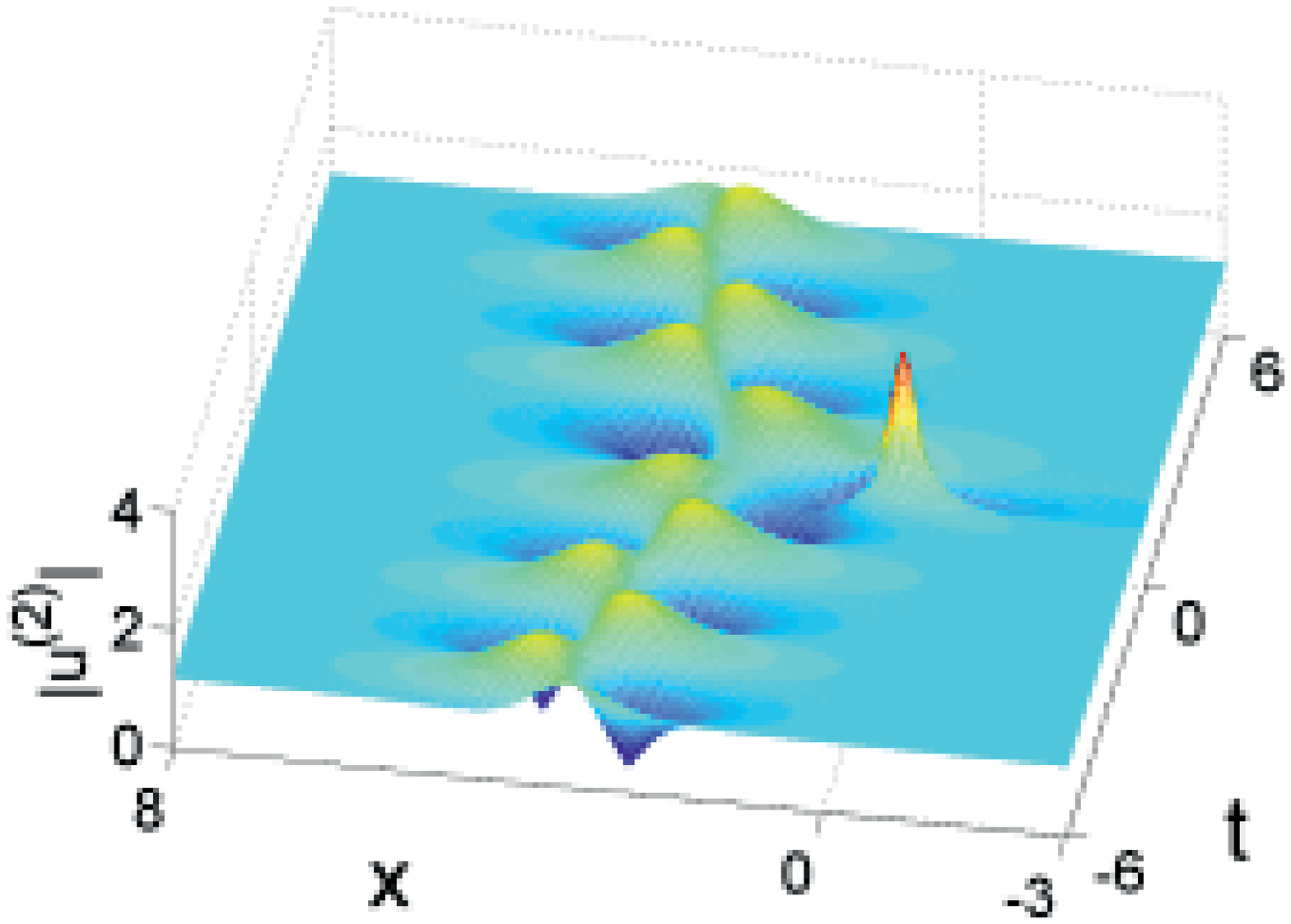}
    \end{center}
     \caption{As in Fig. \ref{fig_peregrine}, with $f=0.1i, a_1=1.2, a_2=1.2$.
    } \label{fig_bre2}
\end{figure}

The superposition of the dark and bright contributions in each of the two wave 
components $|u^{(j)}|$ may give rise to complicated breather--like pulses. These 
results are well represented in Figs. \ref{fig_bdr1}-\ref{fig_bre2}. The single contributions 
of the dark shape $L/B$ and bright shape $M/B$ are better displayed when f.i. $a_2=0$. 
In this case typical distributions $|u^{(1)}(x,t)|, |u^{(2)}(x,t)|$  are displayed in
Figs. \ref{fig_bdr1}, \ref{fig_boom}.
Figure \ref{fig_bdr1} shows a vector dark--bright soliton together with a single Peregrine soliton. 
Decreasing the value of $|f|$, Peregrine and dark--bright solitons separate. By increasing $|f|$, 
Peregrine and dark--bright solitons merge and the Peregrine bump cannot be identified while 
the resulting dark--bright pulse apprears as a boomeron-type  
soliton (see Fig. \ref{fig_boom}), i.e. a soliton solution with a time--dependent velocity
 \cite{degasperis06,conforti06}. Note that the solution (\ref{soliton}) includes as a special case the 
 bright-dark-rogue wave solution that was reported in \cite{guo2011}.

Finally our formula (\ref{soliton}), if all parameters $f, a_1, a_2$ are non vanishing,  
describes the dynamics of a breather--like wave resulting from the interference between 
the dark and bright contributions. Distributions $|u^{(1)}(x,t)|, |u^{(2)}(x,t)|$ which 
are typical of this general case are displayed in Fig. \ref{fig_bre2}.
Again, by decreasing $|f|$ Peregrine and breather solitons separate while 
Peregrine and breather solitons merge, with boomeronic behavior, if instead $|f|$ increases. 

These results provide evidence of an attractive interaction between the dark--bright wave 
and the Peregrine rogue wave. \textit{The observed behavior may also be interpreted as a mechanism of 
generation of one rogue wave out of a slowly moving boomeronic soliton. }
    
Let us discuss the experimental conditions for the observation
of the vector, semi-rational freak solitons. Nonlinear optics is a fertile 
ground to develop the knowledge of the phenomenon of vector freak or rogue waves. 
\begin{figure}[h]
\begin{center}
\includegraphics[width=6cm,height=3cm]{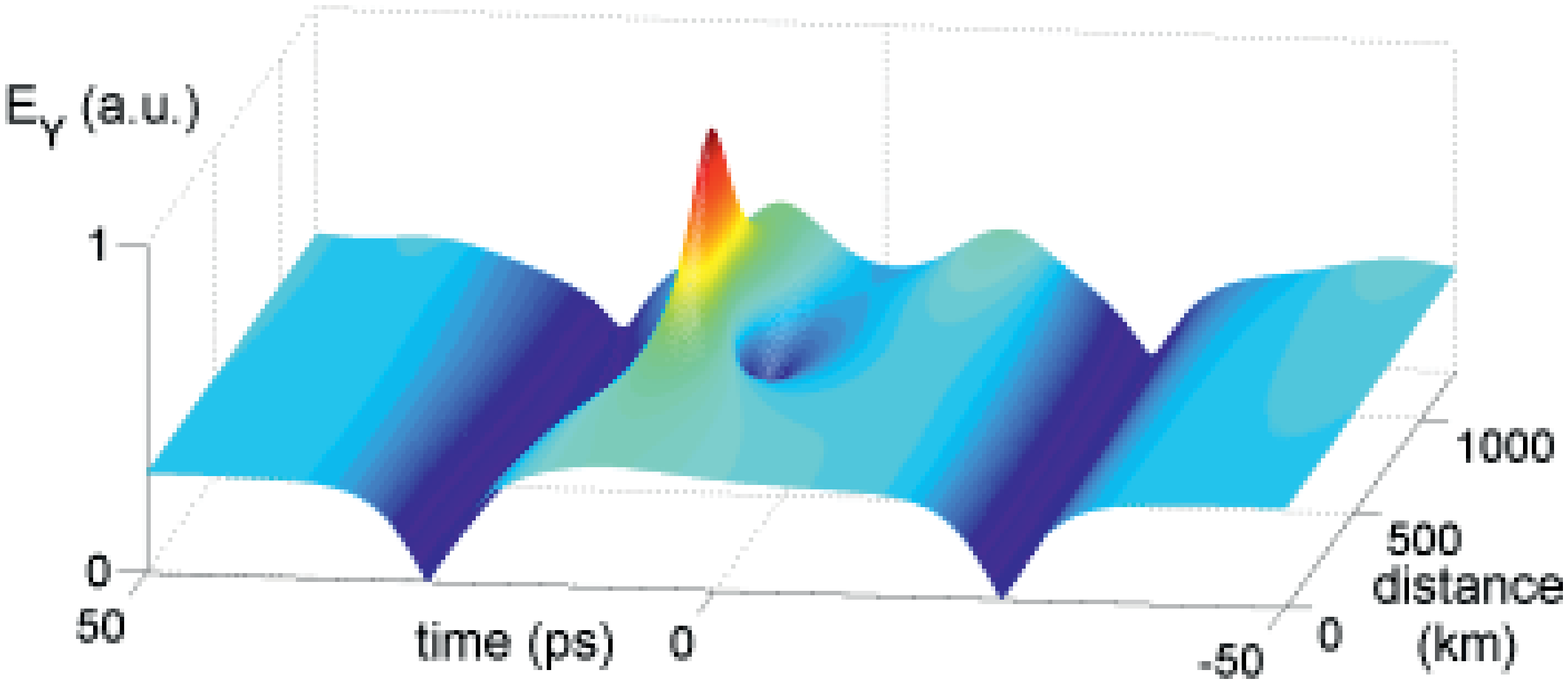}
\includegraphics[width=6cm,height=3cm]{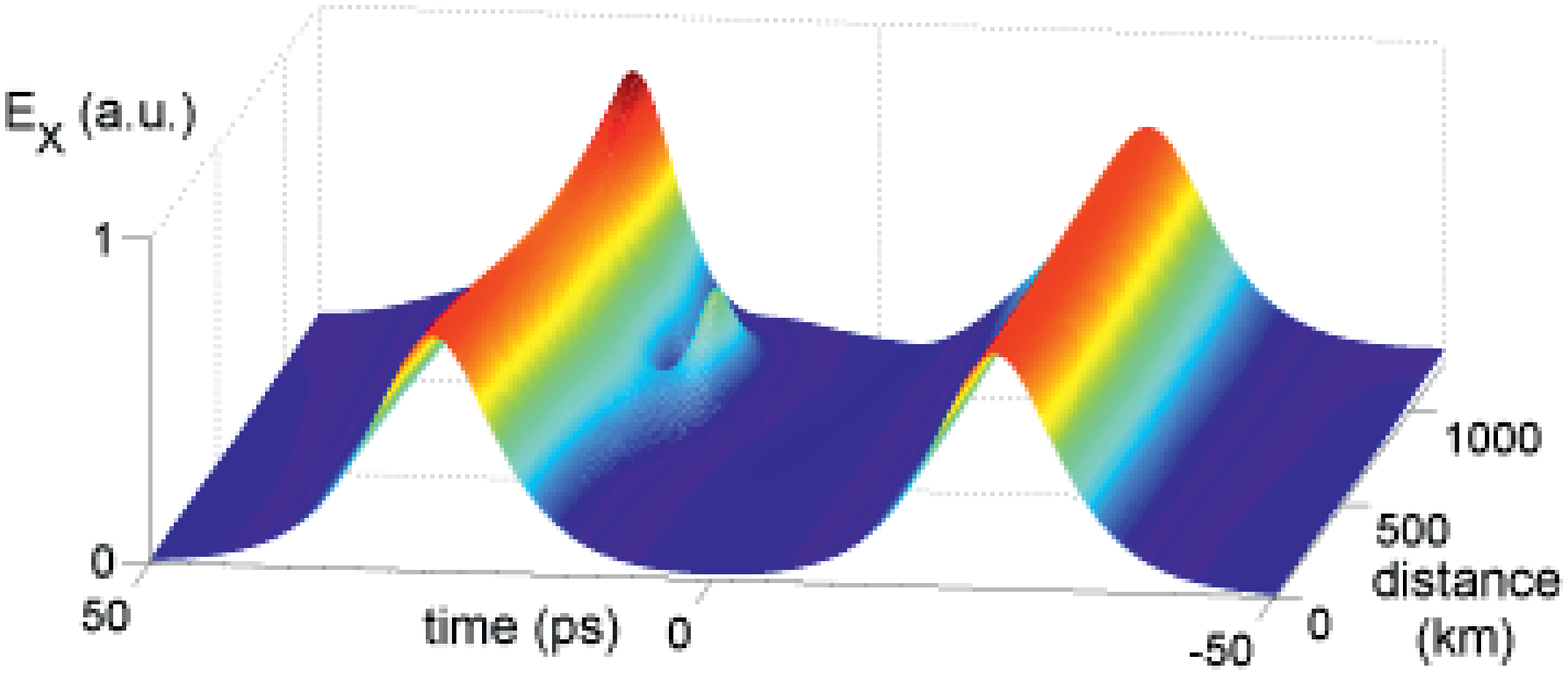}
\end{center}
     \caption{Numerical transmission of two $50$ ps spaced dark-bright solitons 
     in optical fibers, y-polarized dark waves ($E_Y$), and 
     x-polarized bright envelopes ($E_X$).} \label{fig_bd}
\end{figure}
As first scenery, consider the propagation of arbitrarily polarized optical pulses in a 
weakly dispersive and nonlinear dielectric. In fact, Eqs. (\ref{VNLS}) apply to a Kerr 
medium with the electrostrictive mechanism of nonlinearity \cite{Kaplan}, as well as 
to randomly birefringent fiber optic transmission links \cite{evangelides92,wang99}. 
Indeed, the use of the polarization degree of freedom for doubling the capacity of 
long-distance fiber-based transmission systems has been currently widely adopted 
by means of the technique of polarization multiplexing. To be specific, we consider 
the transmission at the $40$ Gbit/s rate of a train of dark-bright solitons, dark in one polarization, 
and bright in the orthogonal polarization.  Fig. \ref{fig_bd} shows that a Peregrine soliton 
is generated at $800$ km, and it attracts a dark-bright soliton. In this example, we numerically
integrated Eqs. (\ref{VNLS}) for properly rescaled wave envelope amplitudes $E_Y$, $E_X$ and rescaled coordinates, with initial conditions two dark-bright solitons plus a small noise seed. We used 
a fiber nonlinear coefficient of $1.3$ km$^{-1}$W$^{-1}$, the anomalous average fiber 
dispersion of $0.1$ ps$^2$/km, and a dark-bright full width at half maximum of $8.25$ ps
($33\%$ of the $25$ ps bit period); the peak power of the two polarizations is equal 
to $3$ mW and $6$ mW, respectively.  
%
%
As second scenery, we may consider incoherently coupled photorefractive 
spatial waves in strontium barium niobate (SBN). Modulation instability and
the existence of unstable dark-bright pairs (first steps in demonstrating 
vector Peregrine waves and dark-bright-Peregrine dynamics) have been already demonstrated in 
SBN \cite{segev97}. Set-ups proposed in Ref. \cite{segev96,segev97} can be exploited to observe
and characterize spatial vector rogue waves in SBN.
%

%

\textit{Conclusions.}\label{sec4}
Here we have analytically constructed and discussed, a multi-parametric vector freak solution
of the vector NLSE. This family of exact solutions includes  known vector Peregrine (rational) solutions,
and novel freak solutions which feature both exponential and rational dependence on coordinates.
Because of the universality of the vector NLSE model (\ref{VNLS}), our solutions contribute to a better control and
understanding of rogue wave phenomena in a variety of complex dynamics, 
ranging from optical communications to Bose-Einstein condensates and financial systems. 


\textit{Acknowledgement. }
The present research was supported in Brescia by the Italian Ministry of University and Research (MIUR) (Project Nb.2009P3K72Z).




\end{document}